\documentclass[prl,aps,twocolumn,notitlepage]{revtex4}

\usepackage{stmaryrd}
\usepackage{color}
\usepackage{mathrsfs}
\usepackage{rotating}
\usepackage{graphicx} 
\usepackage{amsfonts}
\usepackage{amsmath}
\usepackage{mathrsfs}
\usepackage[american]{babel}
\usepackage{slashed}
\usepackage[pdf]{pstricks}
\usepackage{pstricks-add}
\usepackage[off]{auto-pst-pdf}
\usepackage{pst-xkey}
\usepackage{pst-plot}
\usepackage{pst-grad}
\usepackage{pst-eps}

\newcommand{\bea}{\begin{eqnarray}}
\newcommand{\eea}{\end{eqnarray}}
\newcommand{\beq}{\begin{equation}}
\newcommand{\eeq}{\end{equation}}

\begin{document}
\title{Deconfinement transition and black holes}
\author{Antonino Flachi}
%\email{antonino.flachi@ist.utl.pt}
\affiliation{Centro Multidisciplinar de Astrof\'{\i}sica,
Departamento de F\'{\i}sica, Instituto Superior T\'ecnico,
Universidade T\'{e}cnica de Lisboa,\\ Avenida Rovisco Pais 1, 1049-001
Lisboa, Portugal}
%\pacs{03.70.+k,11.30.Rd,11.10.Wx}
%Theory of quantized fields, 03.70.+k
%Chiral symmetries, 11.30.Rd
%Finite-temperature field theory, 11.10.Wx

\begin{abstract}
In this paper we propose an effective description of the transition to deconfinement in the vicinity of a black hole. For this we adapt the approach that uses the dual quark condensate as order parameter for confinement, originally introduced in the context of lattice QCD, to a strongly interacting fermion effective field theory propagating on a curved background. We construct numerically the dual condensate and determine approximately the region of the deconfined phase. The present analysis suggests that quarks will be radiated directly by Hawking emission, while hadrons will form at the boundary of the deconfined region by ``standard'' nonperturbative QCD effects. This example provides a nontrivial setup to discuss how gravity affects confinement. 
\end{abstract}
\maketitle

A correct description of black hole evaporation, even within the semiclassical approximation and ignoring nonequilibrium effects, cannot prescind from a detailed knowledge of particle physics. As predicted by Hawking \cite{hawking}, a black hole of mass $m$ evaporates at a temperature inversely proportional to its mass. In Planck units (used everywhere in the following) $T_{BH} = {1/(8 \pi m)}$. As the evaporation proceeds, temperature increases, and when $T_{BH}$ exceeds the pion mass, first pions and then heavier hadrons can be, in principle, directly produced at the horizon due to Hawking evaporation (see, for example, Ref.~\cite{exmps}). In reality, whether hadrons or elementary partons (from which hadrons will form as QCD jets) are directly radiated is, to the best of our knowledge, an open question whose answer requires a quantitative control over confinement and chiral symmetry breaking in curved space. 

Obvious motivations triggering interest in this problem arise in cosmology, where the evaporation of primordial black holes is used to place constraints on models of the early Universe (see Ref.~\cite{carr} for a review). The same problem is also relevant in the context of TeV-scale gravity, where evaporating miniature black holes are predicted to form in highly energetic particle collisions and possibly observed in cosmic ray facilities or in collider experiments \cite{dimopoulos,giddings}. 
It is clearly prior to any black hole phenomenology to understand how hadrons are produced and whether any difference in the emission rate of composite degrees of freedom may occur with respect to elementary ones.

Essentially the same arguments suggest that critical point sweeping may occur during the evolution of black holes that may serve as probes of the QCD phase diagram. On the one hand, this indicates a less traditional front where to investigate the thermodynamics of QCD (see, for example, Ref.~\cite{ohnishi}); on the other, it raises the question of how the properties of hot and dense QCD change in the presence of a strong gravitational field. 

Motivations of entirely different nature fall under the umbrella of the AdS/CFT correspondence \cite{maldacena,witten}, that relates 
the gravitational dynamics of a $d$-dimensional AdS spacetime to a $(d-1)$-dimensional conformal field theory (CFT), providing a way to connect a classical weakly coupled gravitational theory to a strongly coupled quantum field theory. In this context, one relevant example is connected to the Randall-Sundrum brane model \cite{rs}, where the correspondence had been used to conjecture the impossibility of having large, static, asymptotically flat black holes localized on the brane \cite{emparan,takahiro}. The intuitive reason behind the conjecture is that a classical higher dimensional localized black hole is dual to a 4D black hole plus a strongly coupled conformal field theory. Expecting IR CFT modes to be efficiently radiated would lead to think that the black hole might not relax to a static configuration. Initial numerical computations seemed to support the conjecture (see Ref.~\cite{norihiro} for a review); however, impressive numerical efforts were finally able to show that localized solutions, in fact, exist \cite{figueras,page}. While a clear-cut resolution of this apparent contradiction is not yet available, Ref.~\cite{nemanja} analyzed the situation from the higher dimensional perspective suggesting that additional suppression factors in the emission rate of the CFT sector may occur because of the compositeness of IR CFT states. Despite a suppression that seems to follow from the bulk picture, an effective four-dimensional explanation is still lacking.

Even in the absence of gravity, the problem remains complicated. While our understanding of chiral symmetry breaking is illuminated by the Banks-Casher formula \cite{banks} that relates the eigenvalue density of the Dirac operator with the quark condensate that plays the role of order parameter in the limit of vanishing quark masses, 
the situation for confinement, especially its relation to chiral symmetry breaking, is not as transparent. If it is an obvious statement that quarks should ``know'' about confinement, the transition to deconfinement is only well defined in the heavy quark limit with the Polyakov loop playing the role of order parameter. Away from this limit, no precise criterion to describe the transition to deconfinement is known. QCD lattice simulations at finite temperature, although suggesting the existence of a precise mechanism relating confinement and chiral symmetry, are also not conclusive. While some computations find the same critical temperature for chiral symmetry breaking and deconfinement \cite{bielefeld}, others find a nonvanishing difference \cite{budapest}.

A way to get a handle on the problem in a simplified setting, often used in QCD, is by means of effective field theories, with the Nambu-Jona-Lasinio model being the classic example \cite{nambujona} (see Ref.~\cite{reviews} for reviews). Although chiral models in their original formulation missed entirely any dynamics coming from the Polyakov loop and therefore were not able to describe confinement, attempts to amend this limitation have been discussed in the literature. 
Perhaps the simplest possibility is the model proposed in Ref.~\cite{fukushima2} defined by an effective potential that encodes both the chiral and the Polyakov loop dynamics and that depends on some free parameter (the lattice spacing), tuned to reproduce the critical temperature at the deconfinement transition as obtained, for instance, in lattice simulations. Although in principle possible, extending the model of Ref.~\cite{fukushima2} to curved space is not at all straightforward. 
A different approach that more easily generalizes to curved space can, instead, be realized by relating spectral signatures of the Dirac operator to confinement. Such a possibility has been, for instance, discussed in Refs.~\cite{gattringer,erek} in the context of lattice QCD starting from the observation that color confinement can be described in terms of the center symmetry of the color gauge group \cite{polyakov,susskind,svetisky}. 
Then, an order parameter for the center symmetry (e.g., for confinement) can be engineered by {\it dualizing}~~the quark condensate, i.e., Fourier transforming the chiral condensate generalized with respect to a set of phase-dependent boundary conditions (and with the gauge fields obeying the usual periodic boundary conditions). This transformation converts the quark condensate into the expectation value of an equivalence class of Polyakov loops with winding number $n\in \mathbb{Z}$. Setting $n=1$ singles out a class of loops that changes, under a center transformation, in the same way as the ordinary Polyakov loop and thus offers a new order parameter to describe the transition to deconfinement \cite{erek}. This way of proceeding allows one to discuss confinement by means of functional methods and provides an active platform to study the problem using strongly interacting fermion effective field theories (see Refs.~\cite{kashiwa,ruggieri,mukherjee} for some examples and Ref.~\cite{fischer} for related work). 

While chiral symmetry breaking in curved space has received attention (see Ref.~\cite{sergei} for a review and \cite{ebert,flachitanaka} for a more recent perspective), the only explicit analysis of confinement in curved space we are aware of is that of Ref.~\cite{sasagawa}, where the case of three-dimensional, constant curvature, spherical and hyperbolic spaces is studied. Here, we take a step further in this direction and adapt the approach of Ref.~\cite{erek} to a strongly interacting fermion effective field theory propagating on a generic curved background. Our goal is to provide an approximate description of the transition to deconfinement in the vicinity of a black hole. 

In its simplest form, a strongly interacting fermion effective field theory can be described by the action 
\bea
S = \int d^4x \sqrt{g} \left\{ 
\bar \psi i \gamma^\mu \nabla_\mu \psi 
+ 
{\lambda\over 2N}
\left(\bar \psi \psi\right)^2 
\right\},
\label{action}
\eea
where $\psi$ represents a Dirac spinor with $N$ being the number of quark degrees of freedom, $\lambda$ is the coupling constant, and $g$ is the determinant of the metric, which for a Schwarzschild black hole is
\bea
ds^2&=&f(r) dt^2 +f^{-1}(r) dr^2 +r^2 (d\theta^2 +\sin^2\theta d\varphi^2)~,
\label{geometry}
\eea
where $f(r)=1-2m/r$ and $r_s=2m$ is the horizon radius. 

Following Ref.~\cite{erek}, the transition to deconfinement can be discussed in terms of the dual quark condensate defined as
\bea
\Sigma_n = - \int_0^{2\pi}e^{-\imath \varphi n}\langle \bar\psi \psi \rangle_{\varphi} d\varphi,
\label{dpl}
\eea
where $\langle \bar\psi \psi \rangle_{\varphi}$ represents the generalized quark condensate computed with respect to a set of $U(1)$-valued temporal boundary conditions,
\bea
\psi(x_i,\beta) = e^{-\imath \varphi}\psi(x_i,0).
\label{bcs}
\eea
The inverse temperature is $\beta=1/T$ and $\varphi \in \left[ 0, 2 \pi \right)$. %Setting $\varphi=\pi$ reproduces the standard chiral condensate. 
The generalized quark condensate $\langle \bar\psi \psi \rangle_{\varphi}$ can be obtained by functional minimization of the partition function, which can be computed using standard techniques, essentially adapting the method of Refs.~\cite{flachitanaka,flachirapid} to the present case. Since black hole geometries are not of constant curvature, the chiral and dual condensates will both be spatially varying. In the present analysis, we limit our considerations to the spherically symmetric case $\langle \bar\psi \psi \rangle_{\varphi}= -(\lambda/N) \sigma_\varphi (r)$. Because of the form of the boundary conditions, it advantageous to rescale the metric and cast it in ultrastatic form by a conformal transformation $\hat{g}_{\mu\nu}=g_{\mu\nu}/f$. Then, the standard trick of using the Hubbard-Stratonovich transformation and subsequent integration over the quark fields allows one to express, in the large-$N$ approximation, the partition function, in terms of $\sigma_\varphi (r)$, as
\bea
Z_\varphi &=&
-\int d^{4}x \sqrt{g} \left({\sigma^2_\varphi\over 2\lambda}\right) + \hat{Z}_\varphi + \delta Z_\varphi~,
\label{eff}
\eea
where 
\bea
\hat{Z}_\varphi =
{1\over 2} \sum_{\pm}
\sum_{n=-\infty}^\infty
\mbox{Tr} \ln  \left[
-\hat{\Delta} + \omega_n^2\left(\varphi\right) +
%=\mathscr{A}^{(4)} 
\mathscr{S}_\varphi^{(\pm)}
\right]_{\mu}.
\label{effactvarphi}
\eea
$\hat{Z}_\varphi$, $\hat{R}$ and $\hat{\Delta}$ are, respectively, the partition function, the Ricci scalar and the Laplacian evaluated in the conformally transformed spacetime. The index $\mu$ signifies a renormalization scale on which results explicitly depend due to the nonrenormalizable character of the theory (\ref{action}). For notational convenience, we have defined
%\bea
$\mathscr{S}^{(\pm)}_\varphi := \hat{R}/6 + f \sigma_{\varphi}^2 \pm f^{3/2}\sigma_{\varphi}'$~.
%\eea
The generalized, phase-dependent frequencies are
%\bea
$\omega_n(\varphi) = {2\pi}\left(n+{\varphi/ 2\pi}\right)/\beta$.
%\label{oijphi}
%\\ \Delta &=& {1\over \sqrt{g}} \partial_i \left(\sqrt{g}g^{ij}\partial_j\right)~.
%\eea
The term $\delta Z_\varphi$ compensates for the conformal transformation and can be expressed in terms of heat-kernel coefficients following the general procedure of Ref.~\cite{dowker}, adapted to the present case. A tractable form for (\ref{effactvarphi}) can be obtained in a variety of ways, using, for instance, contour integral techniques or direct heat-kernel methods (see Refs.~\cite{elizalde,toms} for general introductions). Here, we use zeta-function regularization and express (\ref{effactvarphi}) using a Mellin representation for the functional trace. Then, the heat-kernel expansion, that we carry out to fourth order, allows us to obtain the effective action from which the equations for the phase-dependent condensate $\sigma_{\varphi}(r)$ can be explicitly obtained. Numerical approximation is finally used to explicitly construct the solution. Details of the method have been reported in Ref.~\cite{flachidual}. 

\begin{figure}[ht]
\begin{center}
\unitlength=1mm
\unitlength=1mm
\begin{picture}(87,58)
   \includegraphics[height=6.cm]{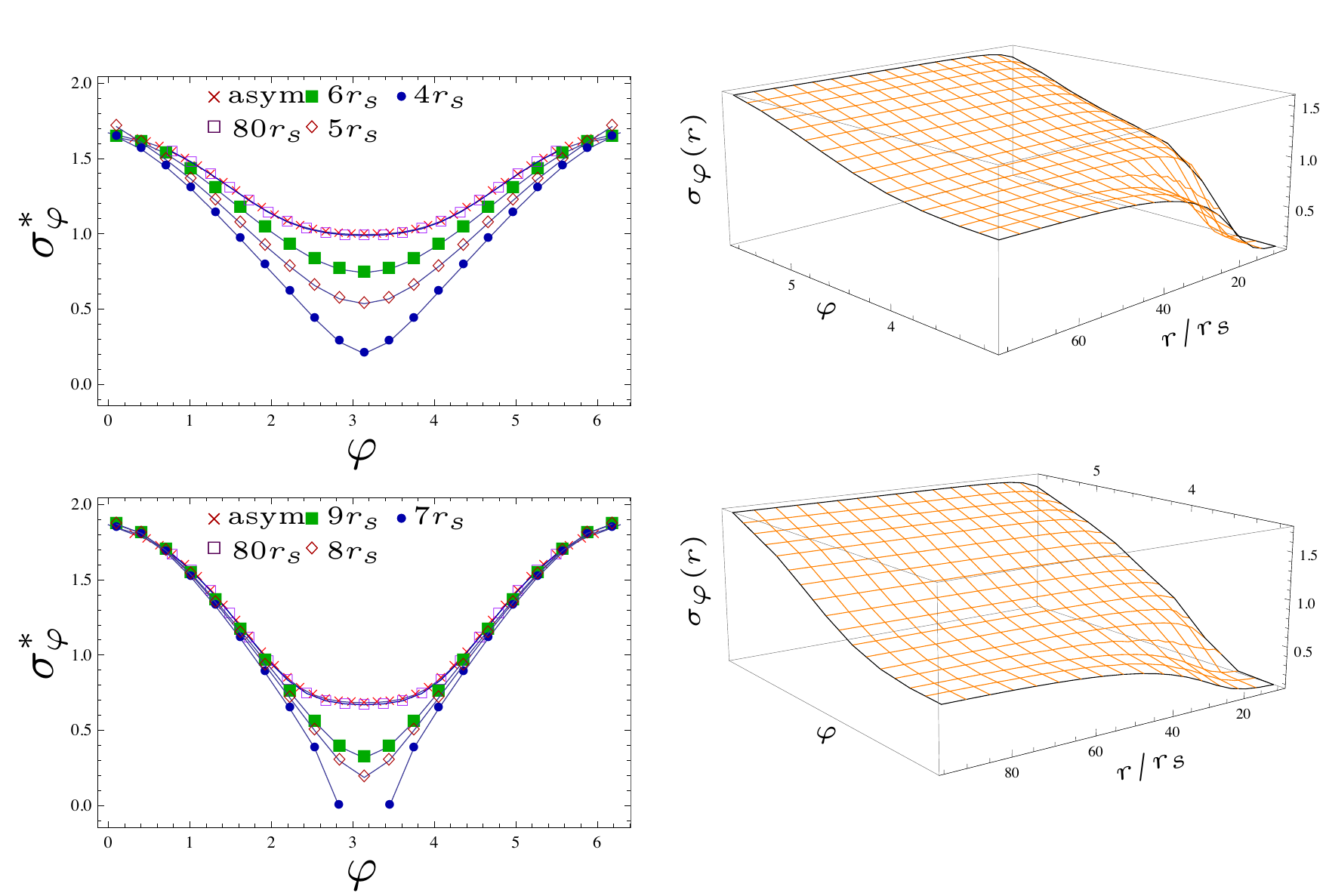}
\end{picture}
\end{center}
\vspace{-.5cm}
\caption{{\it Left:} Dependence of the generalized condensate on the phase $\varphi$, computed by minimizing the effective potential at fixed $r=r_*$, $\sigma_\varphi^*=\sigma_\varphi(r_*)$. The effect of the curvature tends to increase the effective temperature of the system as the horizon is approached as indicated by the curves moving downwards bringing the system into a deconfined phase. The different symbols refer to the value of $r_*$ as indicated. The figures refer to the dimensionless parameter $\sqrt{\lambda} T = 0.20$ (top) $\sqrt{\lambda} T = 0.25$ (bottom). In all figures, we have set the parameters to $\lambda=10^{-2}$, $\mu=10^6$.  {\it Right:} Full three-dimensional generalized condensate $\sigma_\varphi(r)$ obtained by solving the effective equations as described in the text. Only the portion $\pi \leq \varphi \leq 2 \pi$ is shown as the $0 \leq \varphi \leq \pi$ part can be obtained by reflection symmetry. The figures refer to the dimensionless parameter $\sqrt{\lambda} T = 0.20$ (top) and $\sqrt{\lambda} T = 0.25$ (bottom).} 
\label{fig1}
\end{figure}
\begin{figure}[ht]
\begin{center}
\unitlength=1mm
\begin{picture}(150,60)
   \includegraphics[height=6.cm]{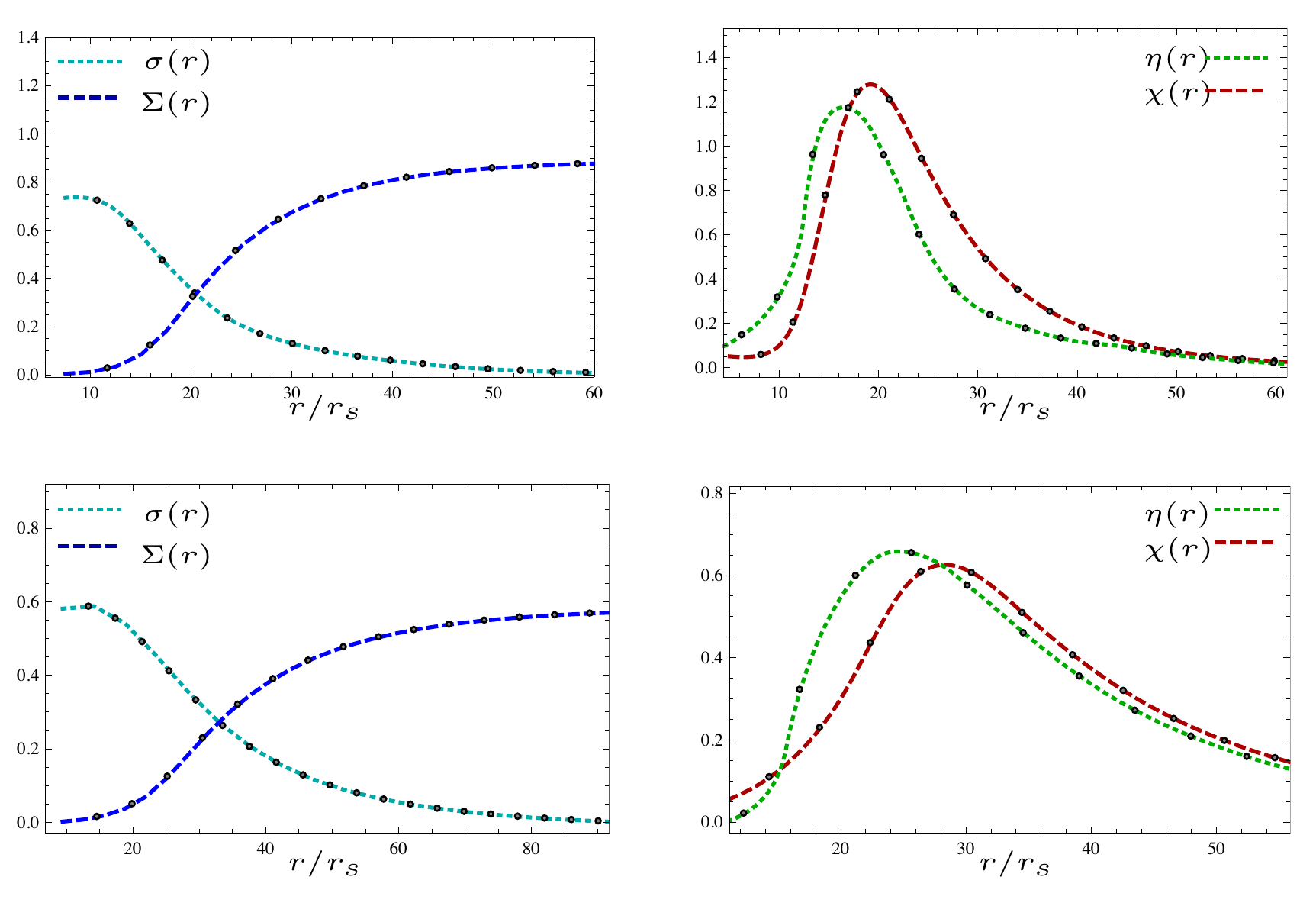}
\end{picture}
\end{center}
\caption{{\it Left:} Chiral (cyan, short-dashed curves) and dual (green dotted curves) quark condensates.
{\it Right:} Chiral (cyan, long-dashed curves) and dual (green short-dashed curves) susceptibilities.
The plots in the top (bottom) panel refer to the dimensionless parameter $\sqrt{\lambda} T = 0.20$ ($\sqrt{\lambda} T = 0.25$).} 
\label{fig2}
\end{figure}

In order to describe how the gravitational field of the black hole affects the transition to deconfinement, we set the coupling constant $\lambda$ and the renormalization scale $\mu$ to obtain asymptotically a phase of broken chiral symmetry since this, as suggested in Refs.~\cite{thooft,casher}, will correspond to a confining phase. 
Here, we use the mass rescaling property of the partition function (see \cite{flachidual}) that allows us to fix the value of $\mu$ arbitrarily. 
Thus, all quantities will be measured in these (arbitrary) units.
Numerical integration is performed by discretizing the phase direction $\varphi$ and using a fourth-order Runge-Kutta scheme to solve for $\sigma_{\varphi}(r)$ at each point of the $\varphi$ grid, $\varphi_i$. The radial coordinate $r$ is taken to vary between $r_s+\epsilon$ and $R$ where $\epsilon$ and $R$ are two fixed cutoffs that we vary in order to check that the solution does not depend on their values. After obtaining the solution $\sigma_{\varphi_i}(r)$ at each point $\varphi_i$, we perform an interpolation along $\varphi_i$ for fixed radial distance by using both the spline and Hermite methods up to third order. The interpolated function is then used to compute the Fourier transform (\ref{dpl}). 

Our results are summarized in Figs.~\ref{fig1} and \ref{fig2}. In the left-hand panel of Fig.~\ref{fig1} we plot the dependence of $\sigma_\varphi$ on the phase $\varphi$ for fixed radial distance and for two indicative values of the black hole temperature. The $\varphi$ dependence is computed by minimization of the thermodynamic potential at fixed $r=r_*$ varying the phase $\varphi$. 
As the horizon is approached, the increase in the gravitational field tends to push the system into a deconfined phase as signalled by the generalized condensate that moves downwards as the horizon is approached.
The full solution of the effective equations for the generalized condensate is shown in the right-hand panel of Fig.~\ref{fig1}.
The numerical solution $\sigma_\varphi(r)$ is then used to compute the dual quark condensate, as defined in (\ref{dpl}), and illustrative results are shown in the left-hand panel of Fig.~\ref{fig2}, where we plotted the chiral [$\sigma(r)=\sigma_{\{\varphi=\pi\}}(r)$, cyan dashed line] and dual quark condensate [$\Sigma(r)=\Sigma_{1}(r)$, green dotted line]. 
We have defined and computed the radial chiral [$\chi(r)$, cyan long-dashed curve] and dual [$\eta(r)$, green short-dashed curve] susceptibilities, $\chi(r) = {\partial \sigma_{\varphi=\pi} / \partial r}$ and $\eta(r) = -{\partial \Sigma^{(1)} /\partial r}$, whose peaks indicate, respectively, the critical distance at which chiral symmetry spontaneously breaks and the transition from a deconfined to a confined phase takes place. The right-hand panel of Fig.~\ref{fig2} illustrates these functions for some values of the parameters.

Simple inspection of Fig.~2 clearly shows that the horizon is encapsulated within a region of deconfined phase separated from the confined region by a smooth crossover (investigation of the nature of the crossover transition in QCD can be found in Ref.~\cite{natfod}). Secondly, the boundary of the deconfined region (defined by the peak in the dual susceptibility) is slightly shifted with respect to that of restored chiral symmetry (defined by the peak in the chiral susceptibility), suggesting that gravitational or geometrical effects may trigger a separation in the critical points at which chiral symmetry breaking and the transition to deconfinement occur.
 
At least in the effective field theory approximation used here, we should expect that elementary partons will be directly radiated by Hawking evaporation occurring at the horizon, while hadrons will form only at the boundary of the deconfined region where QCD jets will form. A sketch is given in Fig.~3.
This suggests that an additional source of suppression in the emission rate of hadrons (to that encoded in the ordinary gray-body factors) should occur due to additional energy necessary for flux tube fragmentation leading to hadron production. A more detailed analysis to see how this can be quantitatively incorporated in the standard Hawking radiation picture is certainly deserved. Also, whether the same idea may be relevant to the discussion of Ref.~\cite{nemanja} should be further investigated. In that context, if correct, this mechanism of suppression in the emission rate of composite modes may be expected to hold whatever the UV completion of the theory might be, as long as it confines. 
\vspace{.0cm}
\begin{figure}[ht]
\begin{center}
\unitlength=1mm
\begin{picture}(80,55)
   \includegraphics[height=5.5cm]{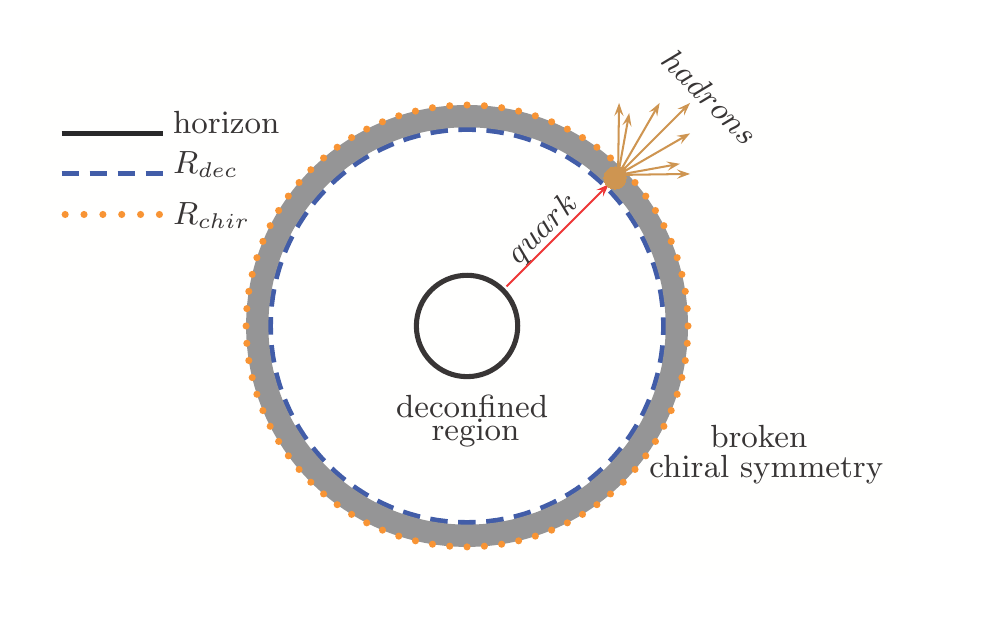}
\end{picture}
\end{center}
\caption{Sketch of possible hadronization process. $R_{chir}$ ($R_{dec}$) indicates the peak in chiral (dual) susceptibility.}
\label{fig3}
\end{figure}

As the black hole mass decreases, the chiral and deconfinement boundaries tend to move away from the horizon. The size of the region of deconfinement can be roughly estimated on dimensional grounds to be, in units of the horizon size, $\alpha T_{BH}/\Lambda_{QCD}$ where $\alpha$ is a constant. While for black holes with mass close to (or larger than) the QCD scale, $\Lambda_{QCD}\sim 200$ MeV, such a region will be very close to the black hole and hadrons will be produced as jets immediately outside the horizon, for cosmologically interesting primordial black holes of mass of approximately $20$ MeV, expected to evaporate at the present epoch, the size of the deconfined region is larger than the black hole horizon radius, possibly leading to a more substantial suppression rate in the hadron emission and generating asymmetries in the hadron distribution due to the shift in the center of production of the jet. 

Two aspects should be more carefully investigated. First, in the computation of the dual condensate we have set, similarly to Ref.~\cite{sasagawa}, the Polyakov loop to zero and the effect of the gauge fields is only included through the effective coupling constant. This induces an error in the dual fermion condensate that can be estimated (even in absence of lattice results in curved space) by computing the general dependence of the dual fermion condensate on the Polyakov loop. Therefore the deconfinement region as computed here should be interpreted as a lower bound on the true value. Including effects of the Polyakov loop will not dramatically change the result, but induces an inaccuracy on $R_{dec}$ making the present results compatible with a coincident peak structure. Secondly, more effort is necessary to achieve a deeper understanding of how gravitational effects may impact on chiral and deconfinement transitions. This can be studied by appropriately analyzing the Ginzburg-Landau expansion for the partition function.

{\it Acknowledgements.} The support of the Funda\c{c}\~{a}o para a  Ci\^{e}ncia e a Tecnologia of Portugal and of the Marie Curie Action COFUND of the European Union Seventh Framework Program (Grant Agreement No. PCOFUND-GA-2009-246542) is gratefully acknowledged. It a pleasure to thank J. Rocha, N. Kaloper and T. Tanaka for their comments on various aspects of this work and J. Rocha for carefully reading the manuscript.

\end{document}